\documentclass[runningheads]{llncs}
\usepackage{graphicx}
\usepackage{booktabs}
\usepackage{verbatim}
\usepackage[tableposition=above]{caption} 
\usepackage{subcaption}
\usepackage{csquotes}
\usepackage{algorithm}
\usepackage{algpseudocode}
\usepackage{xcolor}
\newcommand{\Est}[1]{\hat{#1}}
\usepackage{amsmath,amsfonts}

\usepackage{cite}   
\usepackage[colorinlistoftodos]{todonotes}
\def\notess#1{
\todo[inline, color=green!15]{\sf \small ***SS: #1}}
\def\notei#1{
\todo[inline, color=blue!15]{\sf \small ***IDA: #1}}

\begin{document}

\title{Spatial correlations in the qubit properties of D-Wave 2000Q measured and simulated qubit networks} 
\titlerunning{Spatial correlations in qubit networks}
%
\author{Jessica Park\inst{1,2} \and
Susan Stepney\inst{1}\orcidID{0000-0003-3146-5401} \and
Irene D'Amico\inst{2}
\email{\{jlp567,susan.stepney,irene.damico\}@york.ac.uk}
}
\authorrunning{J. Park, S, Stepney, and I. D'Amico}
%
\institute{Department of Computer Science, University of York, UK  \and
Department of Physics, University of York, UK
}

\maketitle              
\begin{abstract}
We show strong positive spatial correlations in the qubits of a D-Wave 2000Q quantum annealing chip that are connected to qubits outside their own unit cell. By simulating the dynamics of spin networks, we then show that correlation between nodes is affected by a number of factors. The different connectivity of  qubits within the network means that information transfer is not straightforward even when all the qubit-qubit couplings have equal weighting. The similarity between connected nodes is further changed when the couplings' strength is scaled according to the physical length of the connections (here to simulate dipole-dipole interactions). This highlights the importance of understanding the architectural features and potentially unprogrammed interactions/connections that can divert the performance of a quantum system away from the idealised model of identical qubits and couplings across the chip. 

\end{abstract}
\section{Introduction}

Quantum computation is currently being advanced on multiple fronts, including: algorithm development, qubit realisation, device manufacturing, and error correction \cite{Ahn2002-pr, Bharti2022-iu, Harris2009-qi, Pudenz2014-oy}.
Due to the relative infancy and challenging scalability of the technology, the hardware is often hard to control precisely, and the individual qubits can be subject to significant heterogeneity. 
Algorithms will need to be optimised based on the constraints and properties of the hardware, and the hardware will need to be chosen, modified or built based on requirements of the software task. These processes need to be done in parallel such that one the software is not being optimised based on non-optimal hardware and vice versa\cite{Bandic2022-it}.

Different physical realisations of qubits have different levels of robustness to different errors, and so different realisations may be optimal for different functions \cite{Noiri2018-ym,Osada2022-os}.
It seems likely that fabrication inhomogeneities will result in a device where different individual qubits may be optimal for different functions,
potentially allowing improved performance by careful allocation of qubits. 
Before considering how to exploit heterogeneity in the system, it is crucial to understand its sources and effects.
Here we examine how heterogeneity presents itself on a quantum chip, and how this affects the performance when running certain problems.

Section \ref{sec:dwave} gives an overview of quantum annealing and some specifics about the particular architecture that is considered in the remainder of the paper. 
Section \ref{sec:losalamos} presents an investigation in the analysis of spatial correlation that we performed on a dataset provided by Los Alamos National Laboratory \cite{Nelson2022-or}.
The results from this investigation led us to develop and perform tests on a spin network simulator with realistic architectures and dynamics (Section \ref{sec:connection}). Finally, section \ref{sec:Conc} considers the implications of this work and proposes potentially valuable areas of further study.

\section{Quantum Annealing and D-Wave Chimera Architecture}
\label{sec:dwave}

Quantum annealing is a non-universal type of quantum computing most commonly used to find the optimal solution to a problem.
It can do this by finding the global minimum of an energy landscape that encodes the problem. Quantum fluctuation and quantum tunnelling allow the annealer to escape certain local minimal in energy landscapes.

In order to solve such optimisation problems, the cost function (to be minimised) and any associated constraints are formulated into an \textit{Ising Hamiltonian} (modelling the energy of coupled qubits). This is an equation that describes the energy landscape of the system. The desired result of the annealing process is that the system reaches the ground state of this Hamiltonian, which corresponds to the optimal solution of the problem. 

The Hamiltonian that describes quantum annealing is
\begin{equation}
\label{eq:Anneal}
H(x,s) = \frac{A(s)}{2}\left(\sum_{i}\Est{\sigma}_{X}^{(i)}\right)+\frac{B(s)}{2}\left(\sum_{i}h_{i}\Est{\sigma}_{Z}^{(i)}+\sum_{i>j}J_{ij}\Est{\sigma}_{Z}^{(i)}\Est{\sigma}_{Z}^{(j)}\right),
\end{equation}
where $x=\{x_0, x_1, x_i...x_N\}$  is the state of the $N$-qubit system;
$s$ is normalised time;
$\Est{\sigma}_{X}^{(i)}$, $\Est{\sigma}_{Z}^{(i)}$ are the Pauli matrices acting on qubit $x_i$; 
$h_i$ and $J_{ij}$ encode the problem as qubit biases and coupling weights, and, in practice, are limited by the physical hardware graph (qubit-coupling connectivity) of the annealing device. 

Annealing occurs between physical times $t=0$ and $t=t_{f}$, normalised into an annealing fraction: $s={t}/{t_f}$, so $0 \le s \le 1$. 
$A$ and $B$ are functions of  $s$ and their relative magnitudes describe the state of the system as it moves from a general superposition state (the first term) to the solution state (the second term, the Ising Hamiltonian). 

At $t=0$ ($s=0$), the system has $A(0) \gg B(0)$:
the state starts as a general superposition of states. 
The system is  slowly annealed by increasing $B$ and decreasing $A$, until at $t=t_f$ ($s=1$) we have $A(1) \ll B(1)$. This is often referred to as freezing out the quantum fluctuations. At this point the qubits, in an ideal system, are in the ground state of the second term, that is, they are in the state representing the solution to the optimisation problem. 
The annealing process needs to happen slowly enough such that the system does finish in the ground state and not in an excited state of the Ising Hamiltonian \cite{Venegas-Andraca2018-yy}. The point at which $A(s) = B(s)$ is known as the quantum critical point (QCP), by analogy to the theory of phase transitions. 

Eqn.\ref{eq:Anneal} describes an ideal system of perfect qubits and perfect coupling.
Physical devices have limitations, imperfections and inhomogenities, however.
One major limitation of quantum annealers is qubit connectivity:
not all qubit couplings can be realised;
indeed most of the $J_{ij}$ are zero (uncoupled). Another relevant limitation is that even potential couplings can be realised only within a certain range of values and only up to a certain precision. The first restricts the coupling range, and the second is a source of unwanted noise and decoherence. Similar issues affect the qubit biases $h_i$.

\begin{figure}[tp]
	\centering
    \includegraphics[width=0.75\textwidth]{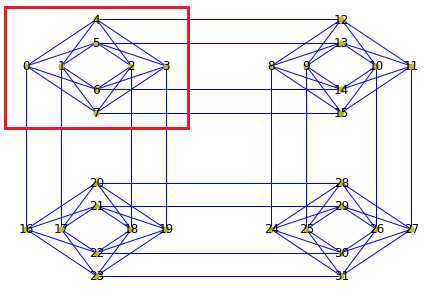}
    \caption{A graph representation of the D-Wave Chimera architecture as present on the 2000Q quantum annealer. The red box shows the 8 qubits that make up a unit cell. (Diagram created using D-Wave NetworkX Python language package \cite{D-Wave_Systems2021-bg}.)}
    \label{fig:Chimera}
\end{figure}

Consider the D-Wave 2000Q, which is the annealer we consider here.
It is designed with 2048 qubits in the `Chimera' architecture, which has 256 unit cells of 8 qubits, arranged in a $16 \times 16$ grid.  
There are connections between qubits inside unit cells and between qubits belonging to different unit cells. Figure \ref{fig:Chimera} shows qubit connections in a $2 \times 2$ grid of unit cells. The yellow dots in the figure represent the qubits; in reality each qubit is an elongated superconducting loop oriented either horizontally or vertically. This and the differences highlighted before may be a source of inhomogeneity in the qubit performances. The full 2000Q chip creates the $16 \times 16$ unit cells by repeating the pattern shown in Figure \ref{fig:Chimera} eight times in either dimension and connecting them in the obvious way.

How a given problem is embedded into this (and other) fixed topologies is the subject of much research.
There is often a requirement to use techniques such as chaining (achieving connections via intermediate qubits) to overcome the limitations \cite{Chancellor2017-xs,Zbinden2020-pw,Raymond2020-qp,Barbosa2021-mr}. This is typically done with an awareness of the overall chip error rate and how that affects the probability of success in practice \cite{Albash2019-zr}. Better characterisation of the individual qubits on the chip would allow for more intelligent and potentially real-time re-configuring embedding algorithms. 

\section{Exploring Spatial Correlations in the Los Alamos data}\label{sec:losalamos}

In order to exploit maximum performance from a given quantum device, 
it is necessary to measure the performance of individual qubits and couplings in that device.
Nelson \textit{et al.} \cite{Nelson2021-ca}
perform repeated sampling of each qubit in their D-Wave 2000Q device through a range of input fields, in a process they refer to as QASA (Quantum Annealing Single-qubit Assessment).
They extract values for four parameters: inverse temperature $\beta$, bias $b$, transverse field gain $\lambda$, and noise $\eta$. 

When this QASA protocol is performed for all the qubits within a chip in parallel, the variations and correlations across the chip (a 16x16 grid of unit cells) can be analysed.
The authors found that the orientation of the qubits (horizontally or vertically aligned superconducting loops) is correlated with both the inverse temperature and transverse field gain parameters. 
They hypothesise that this could be due to \enquote{asymmetry in the chip's hardware layout or to the details of how global annealing control signals are delivered to the qubits} \cite{Nelson2021-ca}.

The Los Alamos National Laboratory (LANL) research group that performed this experiment have made the raw data available,
which we use to perform further investigation into the presence of \textit{spatial correlations} in the four parameters measured for each qubit in the chip, as described in this section. 

To measure spatial 
correlations we use Geary's $C$, a number which determines whether adjacent measurements are correlated \cite{Geary1954-cf}. By adjacent here, we mean qubits that have connections between them, either internal and external to unit cells.
$C$ is defined as:
\begin{equation}
C = \frac{(n-1)\sum_i \sum_j w_{ij}(x_i - x_j)^2}{2 \sum_i (x_i - \bar{x})^2 \sum_i \sum_j w_{ij}},
\label{eq:Gearys}
\end{equation}
where $n$ is the number of qubits,
$x_i$ is the parameter value of qubit $i$,
$\bar{x}$ is the mean value of parameter $x$,
and $w_{ij}$ is the connection weight between qubits $i$ and $j$.  We take $w_{ij} = 1$ for connected qubits,
zero otherwise \cite{Zhou2008-bj}.

$C=1$ represents no correlation, $C=0$ a perfect positive correlation, and $C>1$ an increasingly negative correlation (there is no fixed maximum values for negative correlation).  Positive correlation refers to two variables that tend to move in the same direction. For example, in this case, it would mean that a node with a low bias value tends to be connected to other nodes with low bias values. Negative correlations mean that the value of one node tends to oppose the value of its connected nodes. 

The PySAL package includes a Python script that calculates Geary's $C$, but this could not be used in this case as it requires consecutively numbered nodes \cite{Rey2010-uu}. This data has a number of `dead' qubits in the chip which are not included in the dictionaries of nodes and edges, so their indices are missing.


We calculate Geary's $C$ for the entire dataset (Table~\ref{table:QASA2},  column titled ``all").
The values are very close to 1, indicating little correlation between connected qubits in any of the parameters. This is maybe to be expected if the qubits are well isolated from one another.

We also calculate $C$ for two subsets of the data: involving either just the connections internal to unit cells, or just between unit cells (external). 
The \enquote{all} column represents a weighted average of the \enquote{internal} and \enquote{external} columns; it was calculated using all the connections on the chip, of which there are more internal than external.

Table \ref{table:QASA2} shows that qubits that are connected \textit{between} unit cells show a strong positive correlation in the inverse temperature and transverse field gain parameters, and still rather strong but negative correlations for internal connections within  unit cells. 
Here we label correlations as `strong' when there is more than 10\% difference from the global value found in the \enquote{all} column.

\begin{table}[tp]
    \centering
    \begin{tabular}{l@{\hspace{3mm}}c@{\hspace{3mm}}c@{\hspace{3mm}}c}
           \toprule
     		 & all & internal & external \\
            \midrule
        inverse temperature, $\beta$ 
        & 1.08 & 1.30 & 0.58  \\
        
        bias, $b$ 
         & 0.93  &  0.93 & 0.92 \\
        
        transverse field gain, $\lambda$ 
        & 1.06 & 1.40 & 0.32 \\
        
        noise, $\eta$ 
        &  0.91 &  0.91 &  0.89 \\
        \bottomrule
    \end{tabular}
    \caption{Geary's $C$ spatial auto-correlation of four parameters on the Los Alamos D-Wave 2000Q chip, for all connections, for internal only connections, and for external only connections.}
    \label{table:QASA2}
\end{table}

We might expect that internal connections would correspond to physically closer qubits, and therefore more positively correlated properties, but this does not seem to be the case for these parameters.
We do not actually know the physical distances between qubits in the D-Wave system:
the graphical representation in Figure \ref{fig:Chimera} is just a schematic,
and does not show the real lengths of the different connections. 
When more details on the physical hardware realisation become available, it will be important to confirm if physical separation distance is responsible for the observed correlation between qubits. 


\section{Investigating Different Connection Strengths on Dynamics}\label{sec:connection}

In the LANL QASA experiment, all the connection weights are set to zero, in order to isolate the qubits from any coupling effects. 
Nevertheless, differences are seen in correlations between internal (to the unit cell) and externally coupled qubits,
implying some holdover effect.

Here we investigate correlations explicitly due to coupling strengths that vary due to different coupling lengths.
Due to the planar architecture of the chip, links must be of different physical lengths in order to connect qubits both within and between unit cells. Such physical differences could contribute to differing behaviours 
of qubits.

The spins in a spin network can represent any type of qubit, including the superconducting qubits used in the D-Wave chip. A spin network is a mathematically general model for this purpose.
We have developed a spin network simulator in Python that takes as input a network (based on the Chimera qubit layout shown in fig.\ref{fig:Chimera}) and emulates the natural state dynamics of this is network when one qubit is set to $|1\rangle$ and all other to state $|0\rangle$  at $t=0$. 
The connection weights can be scaled based on relative ratios of their representative lengths in the diagram.  
We test a small simulated network with and without the spin-spin coupling weights having been scaled to their respective lengths.

\subsection{Methodology}
The qubits in the D-Wave chip are physically implemented by rf-SQUIDs (radio frequency Superconducting Quantum-Interference Devices) and the couplings are implemented by Compound Josephson-junction rf-SQUIDs \cite{Harris2009-qi}. 
The way the physical length of a coupling affects its performance is based on the underlying physical processes. 
We chose to investigate repulsive dipole-dipole interactions, which scale with distance as 
\begin{equation}
J \propto \frac{1}{r^3}
\label{dip}
\end{equation}

 to represent the physical interactions taking place within the system.
Eqn. (\ref{dip}) describes well the dominant qubit-qubit interaction for various qubits' physical realisations. Other types of interaction are possible, including interactions beyond nearest neighbours, and will be subject of future investigations.
We compare against a control case where all
coupling weights are equal (corresponding to an $N$-d hypercube layout).


All the coupling strengths are scaled based on the shortest connection having a weight of 1. 
This value is chosen because when the D-Wave chip is operated under normal conditions, all the given coupling weights are rescaled to lie between $-1$ and $1$. We expect the same behaviour for both attractive and repulsive connections so we restrict the experiment to coupling weights between 0 and 1. 


The procedure for defining the Hamiltonian matrix of the simulation is given in Algorithm \ref{alg:HM}. The required inputs define the spin network model as a list of nodes and edges numbered according to figure \ref{fig:Chimera}. The positions (relative coordinates) of the nodes are hard coded into the simulator based on the graphical representation of the chip shown in figure \ref{fig:Chimera}. This section of the code produces an $N \times N$ matrix (the Hamiltonian) where the diagonal terms represent the qubit biases (the $h_i$ values in Eqn.\ref{eq:Anneal}) and the other terms are the coupling weights ($J_{ij}$). If there is an edge connecting nodes $i$ and $j$, then $0 < J_{ij} \le 1$ otherwise $J_{ij} = 0$. The resulting matrix is  symmetric: 
$J_{ij} = J_{ji}$.
In this simulation, we assume all the qubit biases to have the same value, and hence, as the total energy is defined up to a constant, they can be set to zero: $h_i = 0$. 

\begin{algorithm}[tp]
\caption{Create Hamiltonian Matrix($NodeList, EdgeList, ScalingType$)}\label{alg:HM}
\begin{algorithmic}[1]
\State $ds$ := EucLengths(EdgeList) 	\Comment{Distances; Edge lengths are Euclidean distance between the nodes}
\State NodeList, EdgeList := Remap(NodeList, EdgeList) 	\Comment{Remap from native qubit indices to ordered range (0,N)}
\State M := 2D array of size (N, N)	\Comment{Initialise the Hamiltonian matrix}
    \For{idx, item in $EdgeList$}
    
    \If{ScalingFactor = Constant}
        \State $J := J0$
    \ElsIf{ScalingFactor = Dipole}
        \State $J := J0 \cdot (min(ds)/ds[idx])^3$
     \EndIf
     
     \State M[item[0], item[1]] := J
     \State M[item[1], item[0]] := J 
     \EndFor
  \State \text{\textbf{return} M}
  \end{algorithmic}
  \end{algorithm}

The Hamiltonian matrix is used to simulate the time evolution dynamics using a method described e.g. in Mortimer \textit{et al.}\cite{Mortimer2021-ex}. This involves solving Schr\"{o}dinger's equation by expanding $|\Psi(t)\rangle$, the state of the system at any time, in terms of the eigenvectors of the Hamiltonian. This is the preferred method over time step iterations as it doesn't accumulate errors due to the state at each time being calculated directly from the initial state. From the result of this algorithm, the probability of the excitation being measured at each node at each time step can be derived as $|\langle i|\Psi(t)\rangle|^2$, with $|i\rangle$ a shorthand for the state describing the excitation being localised at node $i$. This is referred to as the fidelity of measuring an excitation at a particular node at a particular time. 

\begin{figure}[tp]
\centering
\includegraphics[width=0.75\textwidth]{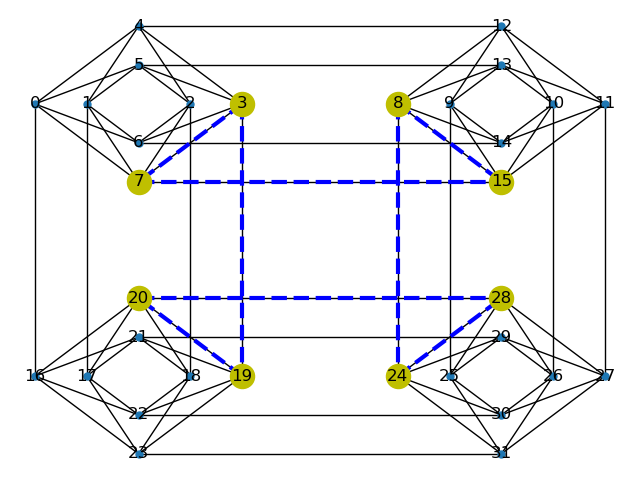}
\caption{Model coded for simulations. The blue connections and the yellow nodes are those used in the simulation. The remaining nodes and connections are shown in black and have been included for completeness.}
 \label{fig:Models}
\end{figure}

A small network with 8 qubits spread over 4 units cells was used for the investigation, this is shown in figure \ref{fig:Models}.

\subsection{Results and Discussion}

In order to consider how the node coupling affects the system dynamics and therefore the spatial correlations in the system, we define a time window within which to consider the information (excitation) transfer through the network. The time window goes from $t=0$ to $t= {1}/{J_{min}}$, where $J_{min}$ refers to the smallest coupling weight in the system.
This time window was chosen because in real quantum devices, the relevant time scales over which operations can be performed is dependent on the strength of the couplings between the qubits. 
The gating time between nearby qubits can be estimated as the inverse of their coupling strength $\sim 1/J$, so $t= {1}/{J_{min}}$ corresponds roughly to the longest gating time in the system, and we can expect the excitation to have propagated through the network by that time. Also, within this time, it is  reasonable to expect that, in  hardware designed for quantum computation, decoherence effects are still extremely low and hence the probability of errors due to additional (and unwanted) interactions remains negligible. Effects of fabrication errors can be taken into consideration within the proposed model, e.g. following Ronke \textit{et. al}\cite{Ronke2011-rp}, however before doing this more information on the hardware details would be desirable.
Within this time window, we consider the excitation fidelity at two specific times: The time at which the first peak in the time window occurs; and the time at which the maximum peak (excluding the initial node) occurs. At these times, the excitation fidelity of all nodes in the system can be measured and compared to infer the correlation between connected nodes. 

\begin{figure}[tp]
\centering
\begin{subfigure}[b]{\textwidth}
            \centering
            \includegraphics[width=0.91\textwidth]{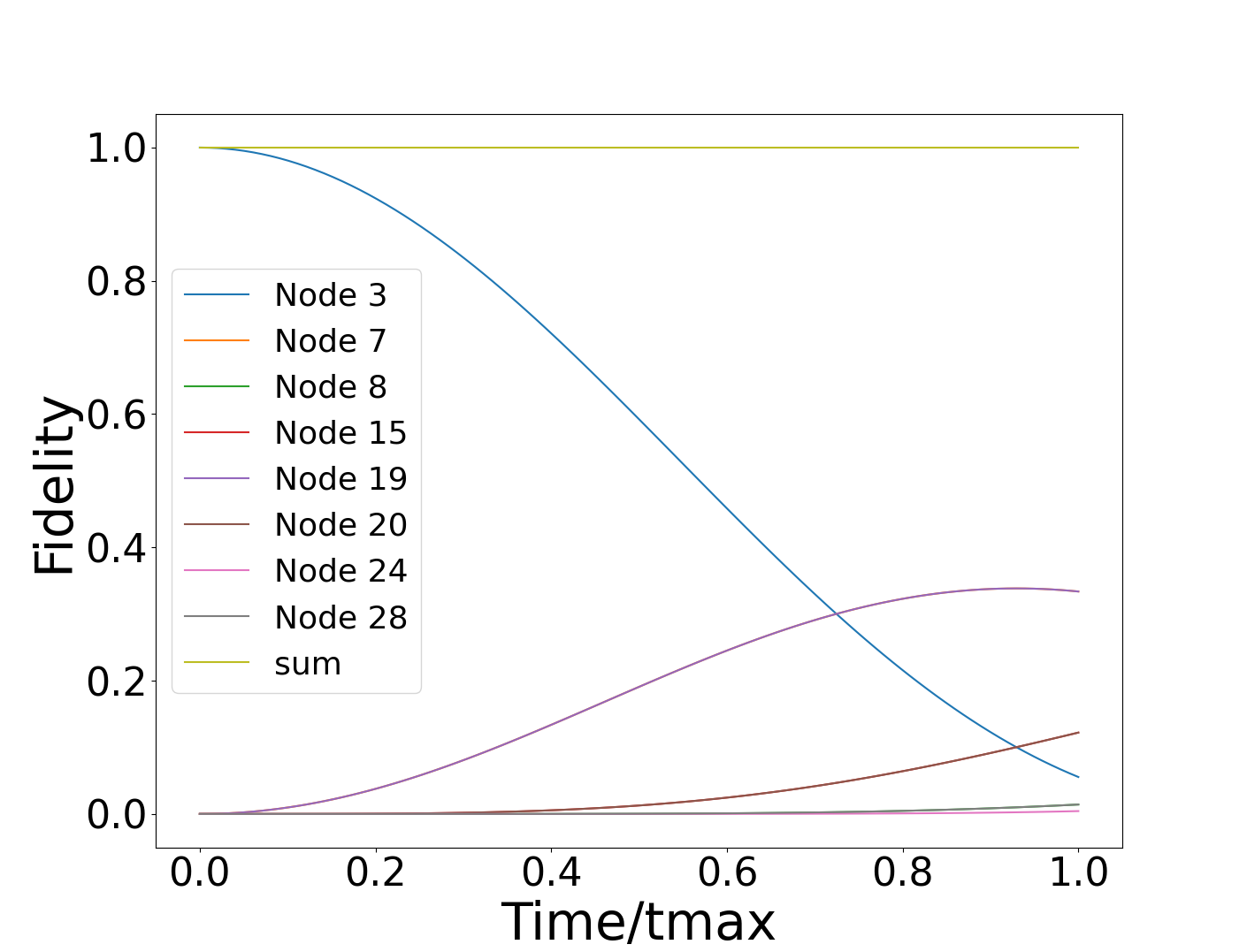}
            \caption{\small 
            coupling weights equal\\\vspace{1em}}
            \end{subfigure}
        \hfill
        \begin{subfigure}[b]{\textwidth}  
            \centering 
            \includegraphics[width=0.91\textwidth]{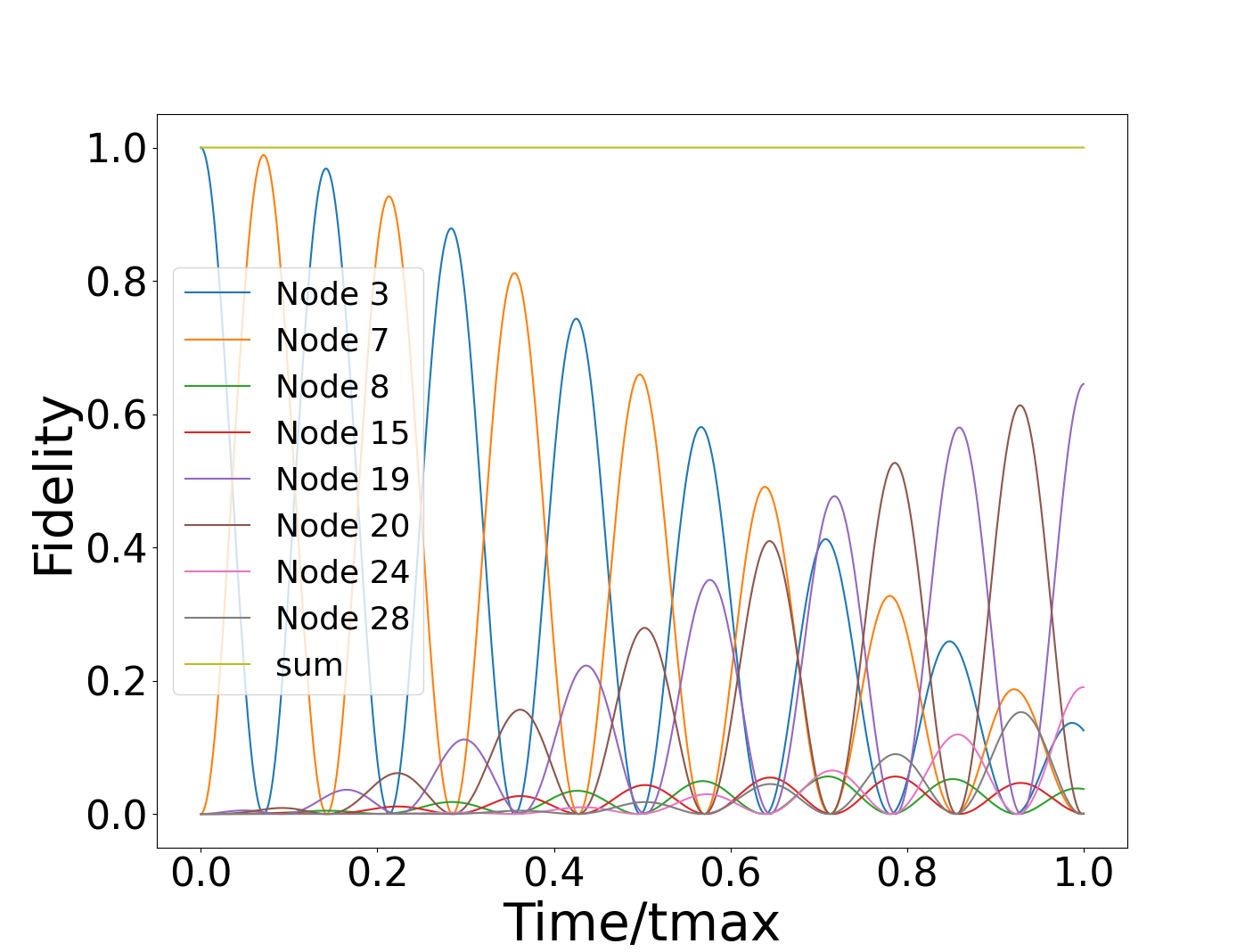}
            \caption{\small 
            coupling weights scaled like dipole-dipole interactions}
            \end{subfigure}
\caption{System dynamics of the 8 node network.}
\label{fig:8N}
\end{figure}

In the 8 node network (Fig.\ref{fig:Models}), each node has one internal and one external connection. All the internal connections are the same length and the external connections are either vertical or horizontal with different lengths. The excitation begins on node 3 at time $t=0$. This node is connected to nodes 7 and 19 with the couplings either weighted equally, or with a dipole-dipole interaction according to their length. 

With constant (length independent) couplings we expect the fidelities of nodes 7 and 19 to have the same dynamics; this is shown in Figure \ref{fig:8N}(a) with the behaviour for node 7 being exactly overlaid by that of node 19.

With dipole-dipole couplings, we expect nodes 7 and 19 to behave differently: the longer external connection here has a coupling strength of only 11\% of that of the shorter internal connection.
So the shorter connection (to node 7) gives rise to larger fidelity peak, and  the longer (to node 19) gives a smaller peak within the considered time-window.
This is a weak enough connection to prevent noticeable peaks in node 19 until approximately $t = 0.2t_{max}$ allowing for a near perfect state transfer between nodes 3 and 7.


The networks have a high degree of symmetry and a cyclic nature which means any excitation transfer to a node could have come via a number of different routes. The 8-node case is topologically equivalent  to a single loop. The excitation could travel around the network both clockwise and anticlockwise passing through each connection once, as well as in any combination of \enquote{backwards} and \enquote{forwards} steps. 
More exactly, since this is a quantum system, the fidelities correspond to the probability of the excitation being measured at the node in question at each time step. The fidelity for each node at each time step includes all the possible routes that the excitation could have taken to be measured at that node.

\begin{figure}[tp]
\begin{subfigure}[b]{0.475\textwidth}
            \centering
            \includegraphics[width=\textwidth]{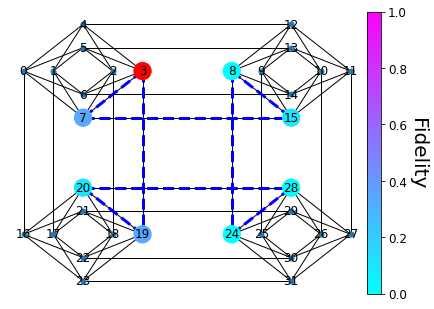}
            \caption[]%
            {{\small Constant coupling at $t=firstPeak$}}
            \label{fig:const_first}
        \end{subfigure}
        \hfill
        \begin{subfigure}[b]{0.475\textwidth}  
            \centering 
            \includegraphics[width=\textwidth]{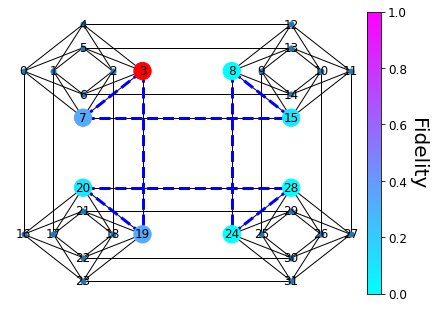}
            \caption[]%
            {{\small Constant coupling at $t=maxPeak$}}    
            \label{fig:const_max}
        \end{subfigure}
        \vskip\baselineskip
        \begin{subfigure}[b]{0.475\textwidth}   
            \centering 
            \includegraphics[width=\textwidth]{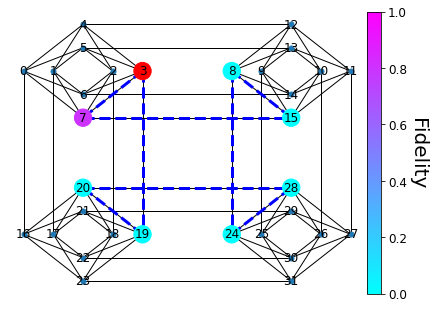}
            \caption[]%
            {{\small Dipole-dipole coupling at $t=firstPeak$}} 
            \label{fig:dip_first}
        \end{subfigure}
        \hfill
        \begin{subfigure}[b]{0.475\textwidth}   
            \centering 
            \includegraphics[width=\textwidth]{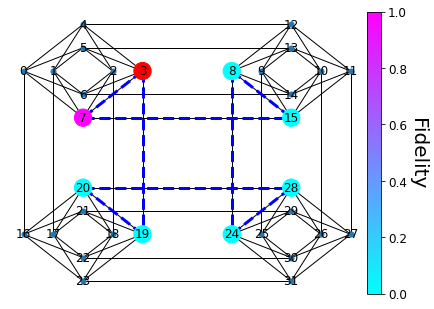}
            \caption[]%
            {{\small Dipole-dipole coupling at $t=maxPeak$}}    
            \label{fig:dip_max}
        \end{subfigure}
    \caption{\small Node fidelities at two different times with two different coupling.}
    \label{fig:8NGraph}
\end{figure}

To investigate  potential spatial correlations, we show the results for the 8N node network, at $t=maxPeak$ and $t=firstPeak$  in figure \ref{fig:8NGraph}. The red node indicates the location of the initial excitation. In the first peak dynamics, it is clear that when the nodes have constant coupling, the edges (3,7) and (3,19) behave identically. When there are dipole-dipole interactions, there is large difference in excitation transfer across these connections with the short connection producing the highest excitation transfer within the observed time window. 

In the constant coupling case, the maximum peak in the time window is also the first peak seen but in the dipole-dipole coupling case the first peak occurs at node 19 whilst the maximum peak is in node 7. At the time of the first peak, node 7 has higher fidelity than node 19 but is still building up and has not yet peaked. This implies that at both peaks, the probability of measuring the excitation at node 7 is higher than at node 19. 

\begin{figure}[tp]
\begin{subfigure}[b]{0.475\textwidth}
            \centering
            \includegraphics[width=\textwidth]{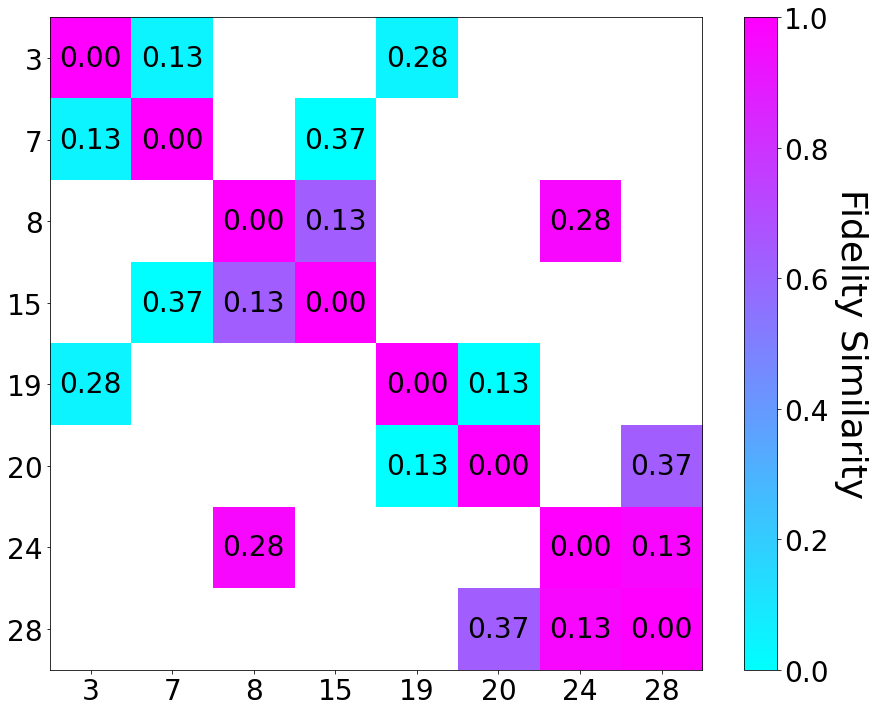}
            \caption[]%
            {{\small Constant coupling at $t=firstPeak$}}
            \label{fig:Gconst_first}
        \end{subfigure}
        \hfill
        \begin{subfigure}[b]{0.475\textwidth}  
            \centering 
            \includegraphics[width=\textwidth]{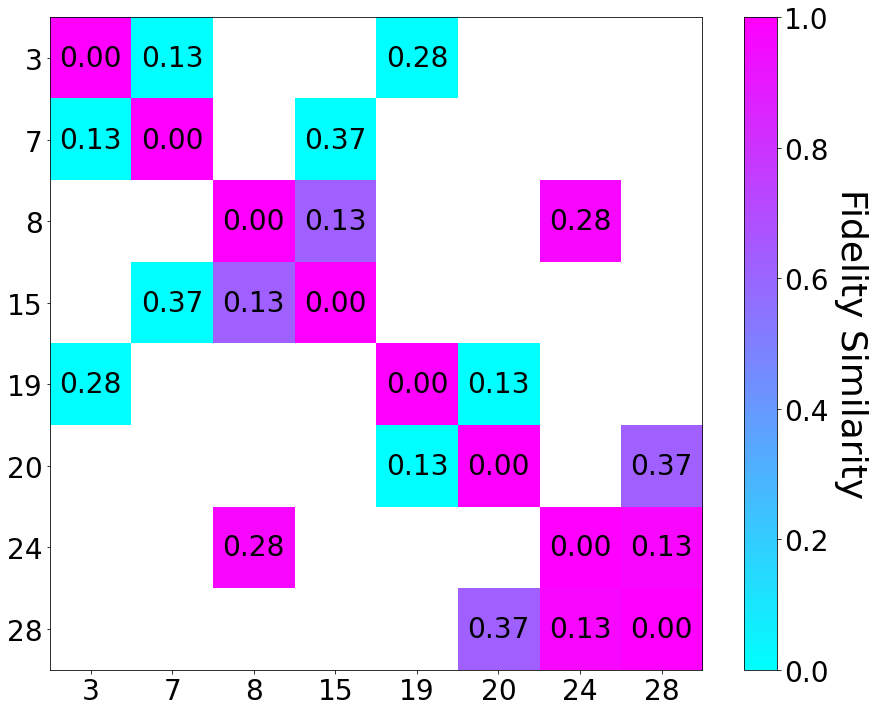}
            \caption[]%
            {{\small Constant coupling at $ t=maxPeak$}} 
             \label{fig:Gconst_max}
        \end{subfigure}
        \vskip\baselineskip
        \begin{subfigure}[b]{0.475\textwidth}   
            \centering 
            \includegraphics[width=\textwidth]{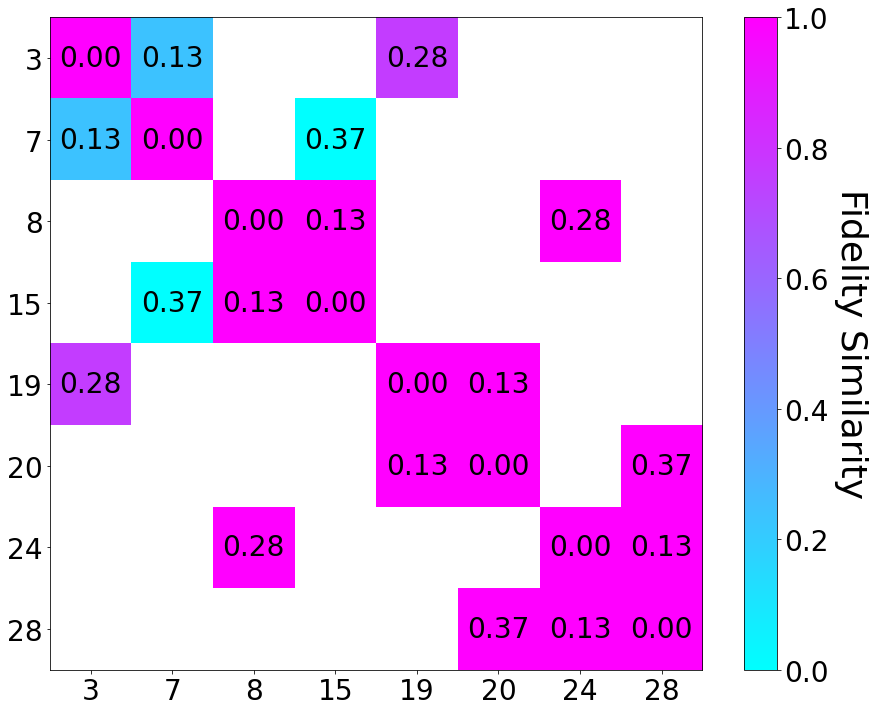}   
            \caption[]%
            {{\small Dipole-dipole coupling at $t=firstPeak$}} 
            \label{fig:Gdip_first}
        \end{subfigure}
        \hfill
        \begin{subfigure}[b]{0.475\textwidth}   
            \centering 
            \includegraphics[width=\textwidth]{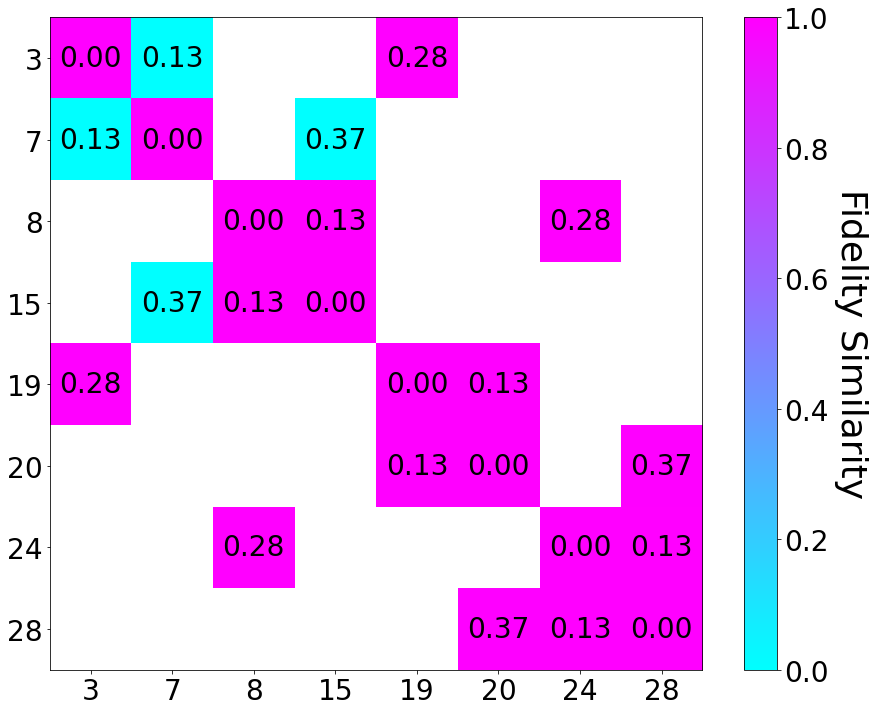}
            \caption[]%
            {{\small Dipole-dipole coupling at $t=maxPeak$}}    
            \label{fig:Gdip_max}
        \end{subfigure}
    \caption{\small  The similarity in the fidelity of connected nodes in the 8 node Chimera architecture. A square at (i,j) is labelled with the relative length of the connection between the nodes i and j and is colourised by the similarity as defined by equation \ref{sim}.}
    \label{fig:8NGrid}
\end{figure}

As well as considering the overall dynamics, to better compare this simulation to the LANL experiments, we also compared the fidelities of all connected nodes. These results are shown in figure \ref{fig:8NGrid}, where each square represents the edge connecting the nodes labelled at its $x$ and $y$ positions. These squares are then coloured by the similarity in the fidelities of the nodes at either end of this edge and are labelled with the normalised connection lengths for reference. 
The similarity is defined here as, 
\begin{equation}
sim=1-|f_{i} - f_{j}|
\label{sim}
\end{equation}
where $f_{i}$ and $f_{j}$ represent the fidelities of the i$^{th}$ and j$^{th}$ node respectively. Therefore if two nodes have similar fidelities, the similarity value is maximum.

At $t=firstPeak$ it is only relevant to consider the top left of the charts as the fidelity of all but the closest 5 nodes from node 3 are all still very close to 0 which means that the similarity between connected nodes is very close to 1. 

In the constant case, we expect the length of the connection to have no effect on the similarity between the connect nodes. Although this is the case in the first row of figure \ref{fig:8NGrid}, in the other rows, this is not the case. We suggest that this is an effect of the fidelity being comprised of the probability of all the different paths that the excitation could have taken from one node to the other. This is a direct effect of the connectivity of the network. 

In the dipole-dipole case, the results are further complicated by the changing connection strengths between the nodes. Intuitively we would expect that a shorter connection length would cause a higher degree of similarity between the nodes. This is not seen at either of the time steps chosen for evaluation here. The charts from the dipole-dipole simulation are noticeably different from the constant case showing that couplings that are affected by physical length will affect the spatial correlations in the system. Because the connectivity is the same in both simulations, the differences must be due to the couplings.

The difference between the constant and dipole-dipole couplings at $t=maxPeak$ is that in the dipole-dipole case, the excitation fidelity is much more concentrated at a small number of nodes meaning that the similarity of these nodes with the others is particularly low. In the constant coupling case, the excitation fidelity is more evenly spread (as seen in Figure \ref{fig:8NGraph}) which means that neighbouring nodes have higher spatial correlation. 

The similarity between connected nodes is affected both by the connectivity of the network as well as the coupling between nodes. This combination gives rise to complex phenomena. This complexity might explain why in the Los Alamos data the longer connections give rise to strong positive correlation. This simulated network only contains 8 nodes, the full D-Wave 2000Q chip contains 2048 qubits (including some `dead' ones). The connectivity in the full chip means that the effects seen in the constant coupling case here would be even more pronounced. 


\section{Conclusions and Future Work}
\label{sec:Conc}

We have shown that there are strong positive spatial correlations in the qubits measured as part of the LANL study on single qubit fidelity beyond the horizontal/vertical delineation shown in the original paper \cite{Nelson2021-ca}. These correlations are only present in the connections between unit cells and not in those internal to unit cells. We hypothesise that this is due to the physical distances between the qubits affecting the connection strengths between them. 
More data, including both from the same device and from other D-Wave 2000Q chips, would be useful in determining whether these correlations seen here are a feature of the particular construction of this kind of chip or even if it is a repeatable phenomenon on exactly same chip. 

To provide evidence for this hypothesis, 
we created a simulated architecture of spins with connection weights that depend on a variable scaling with the connection length, which we compared with the corresponding one having constant coupling strength.  Our results show that even when the couplings between the nodes are independent of length, the dynamics of the system do not behave simply. We suggest that this is due to the connectivity of the network and the multiple paths an excitation could make to transfer from one node to another. Furthermore, when the connection strength is related to the physical distance between qubits, this has significant effects on the dynamics of the system beyond that due to the connectivity. The similarity between nodes that are connected by an edge in the network behaves in a complex (and sometimes counter-intuitive) way that is a combination of the effects due to connectivity and due to physical separation distance (when this is related to coupling strength). 

The differences between the effects of constant and dipole-dipole interaction, combined with the consequences of the connectivity of the network,
highlight the need to understand the effects of specific architectural features over those of the idealised model within the progress of quantum computation. 

The results from our simulations may contribute to explaining the counter-intuitive spatial correlations we found in the Los Alamos data, with longer connections seeming to induce higher correlation. Further investigations into the data presented here would be required to fully understand the causes for the spatial correlation seen in the D-Wave 2000Q chip. 
It would also be beneficial to see how the effects seen here scale up with larger simulated networks, different coupling interactions and longer time periods. These results are to be presented in an upcoming paper. 

Further analysis of a real quantum annealing chip would be able to confirm how closely the effects displayed here affect the dynamics during a quantum annealing cycle. 

\subsection*{Acknowledgements}
The authors wish to acknowledge Defence Science Technical Laboratory (Dstl) who are funding this research.
We thank Carleton Coffrin and his colleagues at the Los Alamos National Laboratory for sharing the data from their Single Qubit Fidelity Assessment.  

Content includes material subject to © Crown copyright (2022), Dstl. This material is licensed under the terms of the Open Government Licence except where otherwise stated. To view this licence, visit http://www.nationalarchives.gov.uk/doc/open-government-licence/version/3 or write to the Information Policy Team, The National Archives, Kew, London TW9 4DU, or email: psi@nationalarchives.gov.uk

\bibliography{paperpile}


\end{document}